\documentclass[sigconf,screen]{acmart}

\usepackage{microtype}

\usepackage{amsmath}
\usepackage{algorithm}
\usepackage{algpseudocode}

\usepackage{graphicx}
\usepackage{booktabs}
\usepackage{tikz}
\usepackage{xcolor}
\usetikzlibrary{positioning,arrows.meta,fit,backgrounds}

\definecolor{pitblue}{HTML}{1F4E79}
\definecolor{pitdark}{HTML}{14324A}
\definecolor{pitlight}{HTML}{EAF2F8}
\definecolor{pitline}{HTML}{B7C9D6}
\definecolor{pitcodebg}{HTML}{F7F9FB}

\setcopyright{cc}
\setcctype{by}
\acmDOI{10.1145/3832783.3834626}
\acmYear{2026}
\copyrightyear{2026}
\acmISBN{979-8-4007-2882-2/2026/10}
\acmConference[ASE '26]{Proceedings of the 41st IEEE/ACM International Conference on Automated Software Engineering}{October 12--16, 2026}{Munich, Germany}
\acmBooktitle{Proceedings of the 41st IEEE/ACM International Conference on Automated Software Engineering (ASE '26), October 12--16, 2026, Munich, Germany}
\acmSubmissionID{ase26tool-p80-p}
\received{2026-05-12}
\received[accepted]{2026-06-19}

\newcommand{\tool}{\textsc{PITMuS}}
\newcommand{\pit}{\textsc{PIT}}

\begin{document}

% -----------------------------------------------------------------------------
% Title and author information
% -----------------------------------------------------------------------------
\title{PITMuS: A Tool for Automated Bug Dataset Generation via Source-Level Mutant Reconstruction}

\author{Tasfia Tasnim}
\orcid{0009-0008-0200-7675}
\affiliation{%
  \institution{University of Texas at Dallas}
  \city{Richardson}
  \country{USA}
}
\email{tasfia.tasnim@utdallas.edu}

\author{Soneya Binta Hossain}
\correspondingauthor
\orcid{0000-0002-7282-061X}
\affiliation{%
  \institution{University of Texas at Dallas}
  \city{Richardson}
  \country{USA}
}
\email{sbhossain@utdallas.edu}

% -----------------------------------------------------------------------------
% Abstract and keywords
% -----------------------------------------------------------------------------
\begin{abstract}
Mutation testing provides a scalable way to generate controlled faults for software testing and empirical software engineering. In Java, \pit\ is a widely used mutation testing tool that creates large numbers of mutants for evaluating test suites. However, \pit\ creates mutants at the bytecode level and reports mutation metadata rather than the corresponding source-level edits. Consequently, these mutants are difficult to inspect, replay, and reuse as structured bug artifacts.

We present \tool, a tool that reconstructs \pit\ mutants at the source level and automatically generates reusable mutation-based bug datasets. Given standard \pit\ reports and the original Java source, and using compiled bytecode when available, \tool\ localizes the affected source statement, applies the corresponding edit, and associates the result with its enclosing method, documentation, and mutation metadata. The tool also supports source-level mutant injection for inspection, replay, and downstream experimentation.

We evaluate \tool\ on eight open-source Java systems. From 69{,}229 mutations reported by \pit, \tool\ produces 69{,}198 source-level original--mutant method pairs across 1{,}913 source files, yielding a source-level pair for 99.96\% of the reported mutation records. These results show that bytecode-level mutation reports can be converted at high coverage into inspectable, context-rich source artifacts for software testing, program repair, and learning-based software engineering.

\smallskip
\smallskip
\noindent\textbf{Screencast:} \url{https://youtu.be/zgHkXnsgciw}\par
\noindent\textbf{Artifact:} \url{https://github.com/assert-lab/PITMuS}
\end{abstract}



\maketitle

\section{Introduction}
\label{sec:introduction}

Many automated and learning-based software engineering techniques benefit from context-rich fault artifacts. Such artifacts support bug localization, program repair, test generation, test-oracle generation and assessment, and documentation-aware automation by associating an original program with a faulty variant while preserving the affected method, documentation, source location, test information, and fault metadata. They allow techniques to be trained and evaluated on explicit source-level changes within their surrounding program context.

Curated benchmarks such as Defects4J~\cite{defects4j} have been indispensable to software engineering research. However, they are necessarily finite and tied to particular systems, versions, and bug collections. Extending them requires substantial manual effort, and they do not automatically evolve as new software versions become available or as research shifts toward new tasks. This limitation is increasingly important in the LLM era because code models are trained on large public corpora, making established benchmarks vulnerable to training-data contamination~\cite{livecodebench,benchmarkleakage,swebenchlive}. A useful complement is therefore to construct fresh, cutoff-aware datasets from selected versions of real software systems by introducing controlled source-level faults.

Mutation testing provides a natural foundation for this purpose. It systematically applies predefined transformations to a program and records whether its test suite detects each resulting mutant~\cite{demillo1978}. In the Java ecosystem, \pit~\cite{pitest} is a widely used mutation testing tool because it is efficient, integrates with common build workflows, and has supported numerous empirical studies~\cite{mutation_survey,hossain2023gaps,hossain2023neural,hossain2025togll}. \pit\ achieves this efficiency by creating mutants at the bytecode level. Its XML reports identify information such as the affected class, method, method descriptor, source line, mutation operator, mutation description, and bytecode index.

\begin{figure}[h]
\centering
\includegraphics[width=\columnwidth]{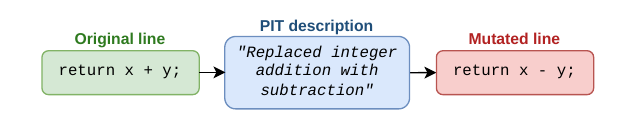}
\caption{A bytecode-level mutation reported by \pit\ and the corresponding source-level edit reconstructed by \tool.}
\Description{An original Java statement returning x plus y is connected to a PIT description that replaces integer addition with subtraction, followed by the reconstructed statement returning x minus y.}
\label{fig:pit-example}
\end{figure}

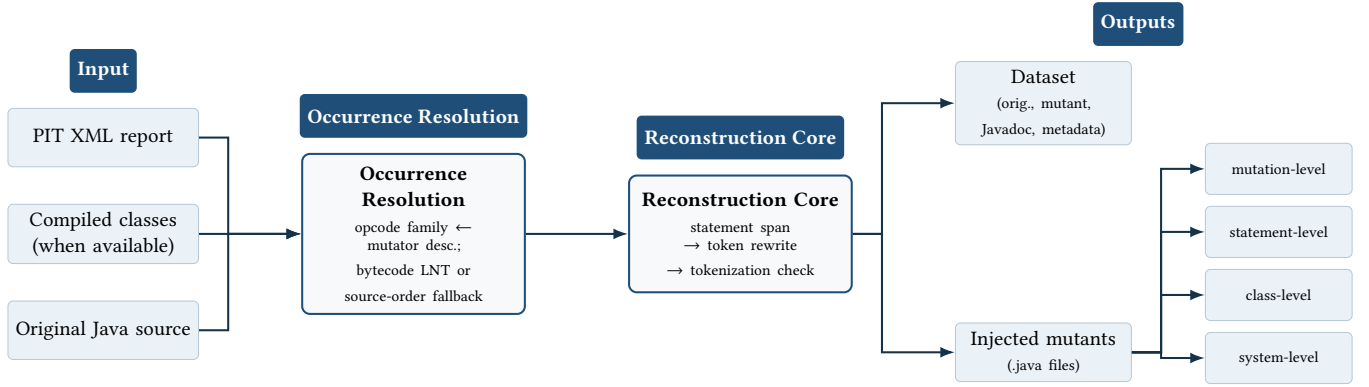
\begin{figure*}[t]
\centering
\resizebox{\textwidth}{!}{%
\begin{tikzpicture}[
    font=\small,
    node distance=8mm and 14mm,
    every node/.style={align=center},
    box/.style={draw, rounded corners=2pt, fill=pitlight, draw=pitline,
                 minimum height=8mm, inner sep=3pt, text width=24mm},
    core/.style={draw, rounded corners=3pt, fill=pitcodebg, draw=pitblue,
                 thick, minimum height=16mm, inner sep=5pt, text width=27mm},
    outbox/.style={draw, rounded corners=2pt, fill=pitlight, draw=pitline,
                    minimum height=7mm, inner sep=3pt, text width=22mm},
    hdr/.style={draw=none, fill=pitblue, text=white, rounded corners=2pt,
                minimum height=6mm, inner sep=3pt, font=\small\bfseries},
    arr/.style={-{Latex[length=2mm]}, thick, draw=pitdark},
]

% --- Inputs ---
\node[box] (xml)  {PIT XML report};
\node[box, below=5mm of xml]   (cls)  {Compiled classes (when available)};
\node[box, below=5mm of cls]   (src)  {Original Java source};

% --- Occurrence resolution ---
\node[core, right=of cls] (res) {\textbf{Occurrence Resolution}\\[3pt]
  {\scriptsize opcode family $\gets$ mutator desc.;\\ bytecode LNT or source-order fallback}};

% --- Reconstruction core ---
\node[core, right=of res] (core) {\textbf{Reconstruction Core}\\[3pt]
  {\scriptsize statement span $\to$ token rewrite\\ $\to$ tokenization check}};

% --- Outputs ---
\node[outbox, above right=4mm and 14mm of core] (dataset) {Dataset\\ {\scriptsize(orig., mutant, Javadoc, metadata)}};
\node[outbox, below right=4mm and 14mm of core] (inject)  {Injected mutants\\ {\scriptsize(.java files)}};

\node[outbox, right=10mm of dataset, yshift=-9mm, text width=18mm] (gran1) {\scriptsize mutation-level};
\node[outbox, below=1.5mm of gran1, text width=18mm] (gran2) {\scriptsize statement-level};
\node[outbox, below=1.5mm of gran2, text width=18mm] (gran3) {\scriptsize class-level};
\node[outbox, below=1.5mm of gran3, text width=18mm] (gran4) {\scriptsize system-level};

% --- Section labels ---
\node[hdr, above=2mm of xml] {Input};
\node[hdr, above=2mm of res] {Occurrence Resolution};
\node[hdr, above=2mm of core] {Reconstruction Core};
\node[hdr, above=2mm of dataset, xshift=13mm] {Outputs};

% --- Arrows ---
\draw[arr] (xml.east) -| ++(4mm,0) |- (res.west);
\draw[arr] (cls.east) -- (res.west);
\draw[arr] (src.east) -| ++(4mm,0) |- (res.west);

\draw[arr] (res) -- (core);

\draw[arr] (core.east) -| ++(4mm,0) |- (dataset.west);
\draw[arr] (core.east) -| ++(4mm,0) |- (inject.west);

\draw[arr] (inject.east) -| ++(4mm,0) |- (gran1.west);
\draw[arr] (inject.east) -| ++(4mm,0) |- (gran2.west);
\draw[arr] (inject.east) -| ++(4mm,0) |- (gran3.west);
\draw[arr] (inject.east) -| ++(4mm,0) |- (gran4.west);

\end{tikzpicture}%
}
\caption{High-level architecture of \tool. Occurrence resolution and statement/method reconstruction are shared between dataset generation and source-level injection, which diverge only at their final output stage.}
\Description{A PIT XML report and original Java source, together with compiled bytecode when available, flow through occurrence resolution and a reconstruction core. The reconstructed edit is emitted either as an original--mutant dataset record or as an injected Java source file at mutation, statement, class, or system granularity.}
\label{fig:architecture}
\end{figure*}

This representation is well suited to mutation-score computation, but it is insufficient for constructing source-level bug artifacts. A \pit\ report indicates that a mutation occurred and describes its general form, but it does not provide the corresponding mutated source code or identify the exact source token or expression that was changed. Consequently, users cannot directly inspect the mutant in an editor, replay it as a concrete source file, or organize it as an aligned original--mutant dataset record with its surrounding method, documentation, and mutation metadata.

Reconstructing the exact source edit is not a simple inverse mapping. A single source line may contain several expressions that match the same mutation description. For example, a line containing both \texttt{index + 1} and \texttt{length + index} contains two candidate additions for a mutant described as ``Replaced integer addition with subtraction.'' Although \pit\ distinguishes these mutations using bytecode-level locations, that information must be reconciled with the original source code to determine which occurrence was changed.

To address this gap, we present \tool, a tool for automated bug-dataset generation through source-level mutant reconstruction. \tool\ combines standard \pit\ XML reports and the original Java source with debug information from compiled class files when available. It uses the mutation description to identify candidate bytecode operations, uses line-number and bytecode-location information to resolve ambiguous occurrences, locates the enclosing source statement and method, and applies an operator-specific source-level rewrite. The reconstructed source is then re-tokenized as a lightweight check for malformed output.

\tool\ supports two complementary uses. First, it injects reconstructed mutants into source files so that users can inspect, replay, and further analyze concrete source-level mutations. Second, it constructs method-level datasets containing aligned original and mutant methods, Javadoc when available, mutation and provenance metadata, and information about covering and killing tests. These capabilities retain \pit\ as the underlying mutation engine while making its mutants available as explicit and reusable source-level artifacts.

This work makes the following contributions:

\begin{itemize}
    \item We introduce \tool, a tool that reconstructs source-level Java mutants from standard \pit\ reports without modifying or replacing the underlying mutation engine.
    
    \item We develop a bytecode-guided reconstruction procedure that resolves ambiguous mutation occurrences, identifies the enclosing source statement and method, applies operator-specific source transformations, and performs a lightweight tokenization check.
    
    \item We provide two utilities: source-level mutant injection and replay, and method-level dataset construction with aligned original and mutant methods, Javadoc when available, mutation metadata, provenance information, and test information.
    
    \item We evaluate \tool\ on eight open-source Java systems. From 69{,}229 mutations reported by \pit, \tool\ produces 69{,}198 source-level original--mutant method pairs across 1{,}913 source files, yielding a pair for 99.96\% of the reported mutation records.
\end{itemize}
\section{\tool\ Workflow}
\label{sec:tool}

The central design principle of \tool\ is to retain \pit\ as the mutation engine while recovering each reported mutation as an explicit source-level edit. \tool\ combines a \pit\ XML report and the original Java source files with compiled \texttt{.class} files when available. The XML report identifies each mutant by its class, method, method descriptor, source line, mutator description, and bytecode index. The compiled class files connect bytecode instructions to source lines, while the source files provide the code that must be localized and rewritten. Figure~\ref{fig:architecture} summarizes this workflow.

\paragraph{Invocation and Requirements.}
\tool\ provides two user-facing commands: \texttt{python gen\_dataset.py <system-path>} for dataset generation and \texttt{python inject.py <system-path>} for source-level mutant injection. The implementation requires Python~3.8+, \texttt{javalang}~\cite{javalang}, and a Java JDK that provides \texttt{javap}. The target system must provide the original Java source files and a \pit\ XML report. Compiled \texttt{.class} files with debug line-number metadata enable precise bytecode-guided disambiguation; when this information is unavailable or inconclusive, \tool\ uses a deterministic source-level fallback.

Both commands use the same reconstruction procedure. Consequently, a dataset record and an injected source file produced for the same mutant are based on the same localized source edit.

\subsection{Shared Source-Level Reconstruction}
\label{subsec:reconstruction}

The principal challenge is that a \pit\ mutation record does not identify the exact source token or expression that was modified. A reported source line may contain several occurrences that are consistent with the same mutation description. \tool\ resolves this ambiguity in four steps.

First, \tool\ maps the textual mutation description to a family of JVM opcodes associated with the corresponding mutation operator. For example, an arithmetic-addition replacement is associated with the typed addition opcodes \texttt{iadd}, \texttt{ladd}, \texttt{fadd}, and \texttt{dadd}; removed-call mutations are associated with the JVM invocation opcodes; and conditional mutations are associated with the relevant comparison opcodes.

Second, \tool\ invokes \texttt{javap -c -p -l} and uses the method descriptor, bytecode index, and line-number table to identify the position of the affected instruction among same-line instructions from the relevant opcode family. This position determines which matching source-level occurrence should be rewritten. If the bytecode information is unavailable or the reported index cannot be resolved, \tool\ falls back to a deterministic left-to-right occurrence counter for mutations sharing the same file, method, descriptor, source line, and description.

Third, \tool\ locates the complete statement containing the reported line. This step is necessary because the affected expression may span multiple lines, as in wrapped conditionals, chained method calls, or multi-line arithmetic expressions. A comment-, string-, and bracket-aware scanner identifies the statement boundaries, while \texttt{javalang} parsing and brace matching locate the enclosing method or constructor.

Finally, \tool\ applies a mutator-specific token-level rewrite. The supported rules cover arithmetic, shift, bitwise, increment, call-removal, return-value, and switch mutations. They also handle conditional-boundary changes, conditional negation and removal, and arithmetic-negation removal. Each rewrite changes only the resolved source span and preserves the surrounding source text and formatting.

If the initial statement-level rewrite fails, \tool\ retries the reported line and then searches a small window of nearby lines. A mutation is retained only when this procedure produces a concrete source change. The reconstructed result is then re-tokenized with \texttt{javalang} as a lightweight validation step. This check can detect malformed output but does not replace project-level compilation or test execution.

Algorithm~\ref{alg:pitmus} summarizes the shared reconstruction procedure.

\begin{algorithm}[t]
\caption{Shared source-level reconstruction in \tool}
\label{alg:pitmus}
\begin{algorithmic}[1]
\Require \pit\ report $R$; source tree $\mathit{Src}$; optional compiled bytecode $\mathit{Bc}$
\Ensure Reconstructed mutant records $\mathcal{Q}$

\State $\mathcal{Q} \gets \emptyset$

\ForAll{$r \in \textsc{ParseXML}(R)$}
    \State $(L,T,\Sigma) \gets \textsc{LoadAndParse}(\mathit{Src}, r)$
    \State $\mathcal{F} \gets
        \textsc{FamilyForDescription}(\textsc{Description}(r))$

    \If{$\mathcal{F} \neq \emptyset$ \textbf{and}
        $\mathit{Bc}(\textsc{Class}(r))$ is available}
        \State $o \gets
        \textsc{ResolveByBytecode}(\mathit{Bc}, r, \mathcal{F})$
    \Else
        \State $o \gets \textsc{NextOccurrence}(r)$
    \EndIf

    \State $(s,e) \gets
        \textsc{StatementSpan}(L,\textsc{Line}(r))$
    \State $u \gets
        \textsc{ApplyWithFallback}(L,T,s,e,r,o)$

    \If{$u=\bot$}
        \State \textbf{continue}
    \EndIf

    \State $(s_m,e_m) \gets
        \textsc{EnclosingMethodSpan}(\Sigma,s)$

    \If{$(s_m,e_m)=\bot$}
        \State \textbf{continue}
    \EndIf

    \State $M \gets L[s_m..e_m]$
    \State $M' \gets M$ with $L[s..e]$ replaced by $u$
    \If{$\textsc{AlreadyEmitted}(M,M')$}
        \State \textbf{continue}
    \EndIf

    \State $\delta \gets \textsc{ExtractJavadoc}(L,s_m)$
    \State $v \gets \textsc{TokenizationCheck}(M')$

    \State $\mathcal{Q} \gets \mathcal{Q} \cup
    \{\langle M,M',\delta,\textsc{Metadata}(r,v)\rangle\}$
\EndFor

\State \Return $\mathcal{Q}$
\end{algorithmic}
\end{algorithm}

\subsection{Automated Dataset Generation}
\label{subsec:dataset-generation}

Dataset generation converts each \pit\ mutation record that yields a distinct source-level edit into a method-level record. For a Java system $P$ with \pit\ report $R_P$, \tool\ constructs

\begin{equation}
\begin{aligned}
\mathcal{D}_P = \{\,&\langle \mathrm{orig}_i,\mathrm{mut}_i,
\mathrm{doc}_i,\mathrm{meta}_i \rangle \mid \\
& r_i \in R_P \land \textsc{Reconstruct}(r_i) \neq \bot\,\}.
\end{aligned}
\end{equation}

Each record represents one reconstructed mutant. Here, $\mathrm{orig}_i$ is the original enclosing method, $\mathrm{mut}_i$ is the same method with the reconstructed source edit, and $\mathrm{doc}_i$ is the preceding Javadoc when available. The metadata $\mathrm{meta}_i$ records provenance information such as the class, method, method descriptor, source file, source line, mutant identifier, bytecode index, mutation description, mutation status, and tokenization-check result. When available in the \pit\ report, it also includes the tests that cover or kill the mutant.

The dataset generator writes the method-level artifacts and Javadoc to \texttt{mutated\_\allowbreak methods.csv} and writes mutation, location, provenance, and test information to \texttt{meta.csv}. The resulting records preserve an explicit correspondence between the original method, its mutated counterpart, and the \pit\ mutation from which the edit was reconstructed.

\begin{table*}[h]
\centering
\caption{\tool-generated datasets across eight Java systems; SLOC denotes source lines of code.}
\label{tab:dataset-stats}
\footnotesize
\renewcommand{\arraystretch}{1.2}
\begin{tabular}{@{}llrrrrrrr@{}}
\toprule
\textbf{Project} & \textbf{Domain} & \textbf{Files} & \textbf{SLOC} & \textbf{PIT Mutants} & \textbf{Pairs} & \textbf{Yield (\%)} & \textbf{Javadoc (\#)} & \textbf{Javadoc (\%)} \\
\midrule
bcel              & Bytecode engineering library & 514 & 41{,}254 & 10{,}786 & 10{,}784 & 99.98  & 8{,}191  & 75.96 \\
commons-beanutils & Reflection and introspection & 258 & 33{,}416 & 4{,}279  & 4{,}279  & 100.00 & 4{,}005  & 93.60 \\
commons-dbutils   & JDBC utility library & 92 & 6{,}863 & 970 & 970 & 100.00 & 815 & 84.02 \\
commons-jexl3     & Java expression language and scripting & 177 & 31{,}659 & 14{,}030 & 14{,}003 & 99.81 & 4{,}908  & 35.05 \\
commons-lang3     & Java language utilities & 401 & 84{,}790 & 17{,}992 & 17{,}992 & 100.00 & 16{,}915 & 94.01 \\
http-request      & HTTP client library & 4 & 4{,}013 & 603 & 603 & 100.00 & 517 & 85.74 \\
joda-time         & Date and time & 330 & 88{,}232 & 13{,}119 & 13{,}117 & 99.98  & 8{,}468  & 64.56 \\
jsoup             & HTML parser and manipulator & 137 & 26{,}125 & 7{,}450 & 7{,}450 & 100.00 & 2{,}094 & 28.11 \\
\midrule
\textbf{Total} & & \textbf{1{,}913} & \textbf{316{,}352} & \textbf{69{,}229} & \textbf{69{,}198} & \textbf{99.96} & \textbf{45{,}913} & \textbf{66.35} \\
\bottomrule
\end{tabular}
\end{table*}

\subsection{Source-Level Mutant Injection}
\label{subsec:source-level-injection}

Source-level injection materializes reconstructed mutants as complete Java source files. For each selected mutant, \tool\ copies the original source file, replaces the resolved statement with the reconstructed statement, and preserves the surrounding code and indentation. Each output file contains one reconstructed mutant, allowing the mutant to be inspected, compiled, tested, or analyzed independently.

\tool\ supports four selection granularities:

\begin{itemize}
    \item \textbf{Mutation-level:} materialize one mutant identified by its mutant ID.
    \item \textbf{Statement-level:} materialize each reconstructed mutant associated with a selected class, method, and source line.
    \item \textbf{Class-level:} materialize each reconstructed mutant reported for a selected Java class or source file.
    \item \textbf{System-level:} materialize every successfully reconstructed mutant in the target system.
\end{itemize}

These granularities determine which mutant records are selected; they do not combine multiple mutations into the same output file. After injection, \tool\ re-tokenizes each generated source file. Files that fail this lightweight check are still written but are marked as invalid so that they can be inspected separately rather than being silently accepted.

\section{Evaluation}
\label{sec:evaluation}

We evaluate \tool\ along three dimensions: (1) the scale of the generated datasets, (2) reconstruction yield, measured as the proportion of \pit\ mutation records that produce distinct source-level original--mutant method pairs, and (3) the records that do not yield such pairs.

\subsection{Subject Systems}
\label{subsec:subjects}

We evaluate \tool\ on eight open-source Java systems listed in Table~\ref{tab:dataset-stats}. Together, they span diverse application domains and project sizes.

\subsection{Dataset Scale}
\label{subsec:dataset-scale}

Table~\ref{tab:dataset-stats} summarizes the datasets generated by \tool. Across the eight subject systems, \tool\ processes 69{,}229 \pit\ mutation records and produces 69{,}198 aligned original--mutant method pairs from 1{,}913 source files, a reconstruction yield of 99.96\%. Among these pairs, 45{,}913 (66.35\%) retain non-empty Javadoc, enabling documentation-aware downstream applications.

\subsection{Reconstruction Yield by Mutation Operator}
\label{subsec:operator-preservation}

\begin{table}[h]
\centering
\caption{Reconstruction yield by \pit\ mutation operator across all eight subject systems.}
\label{tab:per-operator-preservation}
\footnotesize
\renewcommand{\arraystretch}{1.2}
\begin{tabular}{@{}lrrr@{}}
\toprule
\textbf{Operator} & \textbf{PIT Mutants} & \textbf{Pairs} & \textbf{Yield (\%)} \\
\midrule
VoidMethodCall         & 8{,}160  & 8{,}160  & 100.00 \\
NullReturns            & 7{,}152  & 7{,}152  & 100.00 \\
TrueReturns            & 2{,}376  & 2{,}376  & 100.00 \\
FalseReturns           & 1{,}638  & 1{,}638  & 100.00 \\
Increments             & 446      & 446      & 100.00 \\
InvertNegatives        & 80       & 80       & 100.00 \\
PrimitiveReturns       & 3{,}121  & 3{,}120  & 99.97 \\
RemoveConditionals     & 35{,}739 & 35{,}725 & 99.96 \\
ConditionalsBoundary   & 3{,}668  & 3{,}664  & 99.89 \\
EmptyReturns           & 2{,}098  & 2{,}095  & 99.86 \\
Math                   & 4{,}425  & 4{,}417  & 99.82 \\
ExperimentalSwitch     & 326      & 325      & 99.69 \\
\midrule
\textbf{Total}         & \textbf{69{,}229} & \textbf{69{,}198} & \textbf{99.96} \\
\bottomrule
\end{tabular}
\end{table}

Table~\ref{tab:per-operator-preservation} reports reconstruction yield by mutation operator. Six operators achieve a yield of 100\%, and every remaining operator exceeds 99.69\%. Overall, only 31 of the 69{,}229 reported mutation records do not produce distinct source-level original--mutant method pairs.

\subsection{Failure Cases}
\label{subsec:failure-cases}

The 31 records without distinct source-level pairs reflect either bytecode-level duplicates that map to the same source edit and are intentionally skipped, or source-position mismatches that prevent precise localization. Together, they account for fewer than 0.05\% of all reported mutation records.

\section{Conclusion}
\label{sec:conclusion}

We presented \tool, a tool that reconstructs source-level mutants from standard \pit\ reports and automatically generates reusable mutation-based bug datasets. By recovering aligned original--mutant method pairs together with documentation and mutation metadata, \tool\ makes \pit\ mutations easier to reuse in empirical software engineering, software testing, program repair, and learning-based software engineering.

Across eight open-source Java systems, \tool\ produces source-level pairs for 69{,}198 of 69{,}229 reported \pit\ mutation records, a reconstruction yield of 99.96\%. These results show that \pit{}'s bytecode-level mutation reports can be converted at high coverage into inspectable and reusable source-level artifacts without modifying \pit.

\section{Data Availability}

\tool\ is released under the Apache 2.0 License. The source code and demo video are available online~\cite{pitmus}.

\bibliographystyle{ACM-Reference-Format}
\bibliography{main}

\end{document}